\documentclass[twocolumn,english]{revtex4}
\usepackage[T1]{fontenc}
\usepackage[latin9]{inputenc}
\usepackage{graphicx}

\makeatletter
\@ifundefined{textcolor}{}
{%
 \definecolor{BLACK}{gray}{0}
 \definecolor{WHITE}{gray}{1}
 \definecolor{RED}{rgb}{1,0,0}
 \definecolor{GREEN}{rgb}{0,1,0}
 \definecolor{BLUE}{rgb}{0,0,1}
 \definecolor{CYAN}{cmyk}{1,0,0,0}
 \definecolor{MAGENTA}{cmyk}{0,1,0,0}
 \definecolor{YELLOW}{cmyk}{0,0,1,0}
}

\makeatother

\usepackage{babel}
\begin{document}

\title{How does confinement in nano-scale pores change the thermodynamic
properties and the nature of phase transitions of water?}

\author{P.O. Fedichev,}

\affiliation{Quantum Pharmaceuticals Ltd, Ul. Kosmonavta Volkova 6a-1205, 125171,
Moscow, Russian Federation, }

\homepage{http://q-pharm.com}

\email{peter.fedichev@gmail.com}

\author{L.I. Menshikov,}

\affiliation{NRC Kurchatov Institute, Kurchatov Square 1, 123182, Moscow, Russian
Federation}
\begin{abstract}
We analyze thermodynamics of water samples confined in nanopores and
prove that although the freezing temperature can be dramatically lower,
the suppression of the ice nucleation leading to the freezing temperature
depression is a truly macroscopic effect rather than a consequence
of microscopic interactions. The freezing transition itself is a truly
collective phenomenon described by a macroscopic order parameter (the
nearly homogeneous density of the liquid within the pore away from
the pores wall) exactly in the same way as in the bulk liquid. The
thermodynamics properties of the confined and the bulk liquid can
be described by macroscopic thermodynamics and be readily related
to each other simply by proper inclusion of the additional Laplace
pressure exerted by the solid-liquid boundary. 
\end{abstract}
\maketitle
Liquid water confined in nanopores can be prevented from freezing
well below the natural freezing point (see e.g. \cite{jahnert2008melting,morishige1999freezing,schreiber2001melting,petrov2009nmr,mazza2009cluster}
and references therein). The appearance of the liquid water at such
extremely low temperatures was often used to probe the properties
of supercooled water experimentally. More recently the dielectric
properties of the water samples in MCM-41 nanopores were studied within
the temperature range corresponding to the $\lambda-$point of the
bulk liquid \cite{bordonskiy2012study,fedichev2011experimental}.
The observed $\lambda$-feature of the dielectric constant within
the liquid phase of the sample is a signature of a a ferroelectric
phase and thus seem to confirm earlier theoretical predictions made
for a bulk supercooled water \cite{men2011possible}. The suggested
relation brings up a natural criticism since there is no clear cut
answer to the question up to a which point and how the properties
of the confined water samples can be related to the properties of
the bulk liquid? In particular how similar or how different are the
phase transitions and thermodynamics properties of the confined and
bulk liquid? 

The question has a long history and a good summary on the existing
views can be found in \cite{stanley2010water}. Indeed, once confined
within a nano-scale pore a water sample does differ from a bulk liquid
in a number of important ways: 

1) the freezing temperature is dramatically lower (the so called freezing
temperature depression phenomenon) and can be made as low as $-100^{0}C$; 

2) the cooling/freezing cycle of the confined water shows hysteresis,
i.e. the difference between the melting and the freezing temperatures
in a single experiment as described in e.g. \cite{jahnert2008melting,morishige1999freezing,schreiber2001melting,petrov2009nmr,bordonskiy2012study,findenegg2008freezing};

3) on top of that, the neutron scattering experiments combined with
a number of Molecular Dynamics (MD) simulations \cite{steytler1985neutron,bellissent1993structural,takamuku1997thermal,mancinelli2011structural,ricci2008similarities,bruni1998water,awschalom1987supercooled,brovchenko2008multiple,mancinelli2010controversial,mazza2009cluster}
show that the water molecules show complex distribution patterns and
the (molecular) density of liquid within the confined water samples
is far from being homogeneous, which is characteristics of a bulk
liquid.

On the other hand, as also highlighted in \cite{stanley2010water},
the MD simulations of water interacting with a hydrophilic surfaces
of e.g. silica-based nanopores such as MCM-41 show that the hydrophilic
surfaces do not have serious effects on the properties of water, except
for significantly lowering the freezing temperature and stabilizing
the liquid phase \cite{giovambattista2006effect,gallo2010dynamic}.
Similarly only negligible hysteresis was observed in the MCM-41 confined
system by means of X-ray scattering and calorimetric experiment \cite{morishige1997x,schreiber2001melting}.
In another example, the difference between melting and freezing temperatures
is only $\triangle T_{H}\sim1K$ in MCM-41 $3.5nm$-samples \cite{petrov2009nmr}.

Here we revisit the arguments and show that in spite of the apparent
differences the properties of the water samples confined in nanopores
are extremely relevant for the comprehension of the low-temperature
properties of the bulk water, especially if collective properties
of the water samples are probed. In spite of the fact that the freezing
temperature of the small samples can be made dramatically lower, the
suppression of the ice nucleation leading to the freezing temperature
depression is a truly macroscopic effect rather than a consequence
of microscopic interactions. The freezing transition itself is a truly
collective phenomenon described by a macroscopic order parameter characterizing
the nearly homogeneous density of the liquid within the pore away
from the pores wall. The thermodynamics properties of the confined
and the bulk liquids are still macroscopic and can be immediately
related to each other simply by proper inclusion of the additional
Laplace pressure arising from the interaction with the pores walls. 

\begin{figure}
\includegraphics[width=0.9\columnwidth]{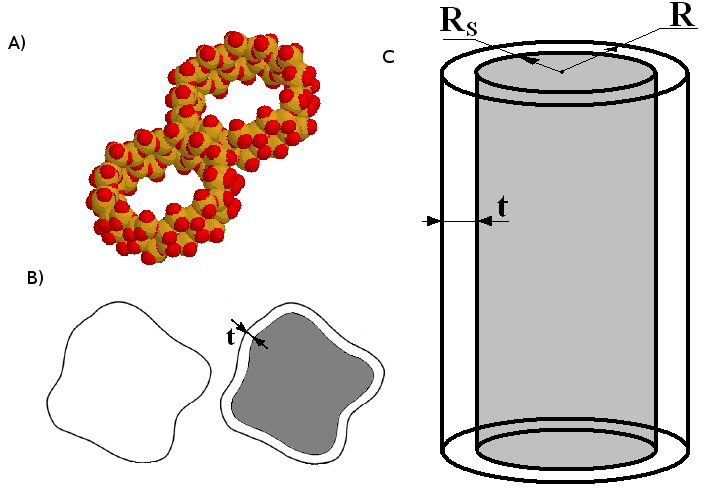}

\caption{The 3D representation of the two cylindrical pores of MCM-41 (A).
The schematic representation of a cross-section (B) of a water sample
confined to a pore at two the temperatures above (left) and below
the freezing point (right). The ice phase region is shown by gray,
non-freezing layer is next to the pore wall. The idealized model (C)
of a cylindrical pore (see the explanation in the text). \label{fig:Pore before and after freezing}}
\end{figure}

To make the point let us carefully derive the freezing temperature
depression $\triangle T_{f}$ using the macroscopic arguments only.
Typical experimental system, e.g. MCM-41 (Mobile Crystalline Material)
is a silicate obtained by a templating mechanism \cite{kresge1992ordered}.
By changing the length of the template molecule, the width of the
channels can be controlled to be within 2 to 10 nm. The walls of the
channels are amorphous $SiO_{2}$ (see Figure \ref{fig:Pore before and after freezing}A).
To simplify the analysis we follow \cite{petrov2009nmr,faivre1999phase}
and consider a simple cylindrical pore of the length $L$ and radius
$R$ filled by liquid water at sufficiently high temperature above
the freezing transition first. As the temperature decreases there
will be a growing body of ice within the pore. To model the appearance
of the so-called ``non-freezing layer'' \cite{petrov2009nmr,schreiber2001melting,morishige1999freezing}
let us assume the ice filling the coaxial cylinder of radius $R_{S}=R-t$,
where $t\approx0.4nm$ is the thickness of a narrow liquid layer next
to the cylinder walls that does not freeze (see Figure \ref{fig:Pore before and after freezing}B,
C). Without loss of generality we further neglect the difference between
the particle densities of the liquid and the solid states and assume
$n_{S}=n_{L}$. Then at every specified value of pressure $P$ each
of the states of the liquid can be characterized by the corresponding
chemical potentials (i.e. Gibbs energy per molecule), $\mu_{L}\left(P,T\right)\equiv\mu_{L}$
and $\mu_{S}\left(P,T\right)\equiv\mu_{S}$, respectively. 

The liquid interacts with the ice body forming within the pore, which
leads to the additional Laplace pressure, associated with the interface
curvature: $\triangle P_{L}=\gamma_{SL}/R_{S}$, where $\gamma_{SL}=32erg/cm^{2}$
is the surface tension coefficient characterizing the ice-liquid boundary
\cite{hillig1998measurement}. Therefore the ice is additionally compressed
by the Laplace pressure $P+\triangle P_{L}$. As the ice grows and
the ice sample radius increases by $dR_{S}$, then the number of molecules
in the solid phase increases by $dN=n_{S}\cdot2\pi R_{S}dR_{S}$ molecules
transferring from the liquid to the ice. Consequently the Gibbs energy
of the system changes by $dG=dG_{S}+dG_{L}$, where $dG_{S}=\mu_{S}\left(P+\triangle P_{L},T\right)dN$,
and $dG_{L}=-dN\mu_{L}\left(P,T\right)$ are the Gibbs energy changes
in each of the phases. Using Gibbs-Duhem relation, $d\mu_{S}=dP/n_{S}-s_{S}dT$,
where $s_{S}$ is the entropy per one molecule in the solid state,
we find that 
\[
dG_{S}\approx\left[\mu_{S}\left(P,T\right)+\delta\mu_{S}\right]dN,
\]
where $\delta\mu_{S}=\triangle P_{L}/n_{S}$. Accordingly, the Gibbs
energy of the ice sample formation can be obtained after integration
of $dG$ from $R_{S}=0$ to its finite (current) value $R_{S}$: 
\[
\triangle G\left(R_{S}\right)=G\left(R_{S}\right)-G\left(0\right)=\triangle\mu\cdot n_{S}\pi R_{S}^{2}L+2\pi R_{S}L\gamma_{SL},
\]
where $\triangle\mu=\mu_{S}-\mu_{L}$. Using the equilibrium condition
$\triangle G=0$ at $R_{S}=R-t$ we can establish the equilibrium
freezing temperature, $T_{f}=T_{\infty}+\triangle T_{f}$, where $T_{\infty}$
is the freezing temperature of the bulk liquid satisfying the equation
$\triangle\mu_{\infty}=\mu_{S}\left(P,T_{\infty}\right)-\mu_{L}\left(P,T_{\infty}\right)=0$.
Since the freezing occurs at $P={\rm const}$, $\triangle\mu\approx\triangle\mu_{\infty}+\triangle T_{f}\left(s_{L}-s_{S}\right)=\triangle T_{f}\triangle h/T_{\infty}$,
where $\triangle h$ is the fusion enthalpy per one molecule and hence
\begin{equation}
\triangle T_{f}=-C/\left(R-t\right),\label{eq:DeltaTfreezing}
\end{equation}
where $C=2\gamma_{SL}T_{\infty}/\triangle H=52.4nm\cdot K$, and $\triangle H=n_{S}\triangle h=3.3\cdot10^{9}erg/cm^{3}$.
The result agrees well with the experimentally established  values:
$C=52.4\pm0.6nm\cdot K$ \cite{jahnert2008melting}, $C=52\pm2nm\cdot K$
\cite{schreiber2001melting} (see \cite{petrov2009nmr} for a more
complete list of experiments).

The very fact that the freezing temperature depression (\ref{eq:DeltaTfreezing})
can be reliably derived from the macroscopic thermodynamics arguments
only is a good indication of the macroscopic nature of the freezing
transition and the freezing depression phenomenon in water confined
in rather wide pores ($R\gg t$) even at a very low temperature. The
derivation shows that the freezing temperature depression (\ref{eq:DeltaTfreezing})
can be simply explained by the increase of the Gibbs energy of the
ice by $\triangle G_{S}=V_{S}\triangle P_{L}=N_{S}\delta\mu_{S}$
due to the additional Laplace pressure $\Delta P_{L}$ exerted by
the solid-liquid interface curvature.

Further analysis of the proposed derivation lets us to draw or to
restate a number of important conclusions. First, the hysteresis phenomenon
brought up in \cite{stanley2010water} to highlight the difference
between the confined samples and the macroscopic liquid can be well
explained within macroscopic thermodynamics as a consequence of the
metastability of the liquid and solid phases in the course of a supercooling
and/or super-heating experiments \cite{shimoda1986nmr,morishige1999freezing,petrov2009nmr}.
As in a ferromagnet, the difference between the freezing and the melting
temperatures (the measure of the hysteresis) $\triangle T_{H}$ is
determined by the value of the free energy barrier $\triangle G_{B}$
separating the molten and frozen states of the pore and as such depends
on the specific path taken to overcome the barrier. According to our
derivation the barrier height, $\triangle G_{B}=\pi\gamma_{SL}L\left(R-t\right)/2$,
corresponds to the Gibbs energy of the half-frozen state of $R_{S}=\left(R-t\right)/2$.
Note that because of the linear dependence of the barrier energy $\Delta G_{B}$
on the sample size $L$, in actual experiments the ice shows up in
smaller pieces characterized by smaller values of $L$ first.

Second, the freezing temperature of the whole sample is defined by
the single free energy balance, $\triangle G=0$. The molecular density
of the water samples is practically homogeneous at $R\gg t$ and hence
freezing occurs everywhere within the water sample at the same time
at a single temperature. Therefore the freezing transition is a collective,
cooperative phenomenon. This picture is supported by the molecular
dynamics \cite{gallo2010dynamic} simulations and means the water
in sufficiently large pores is to a good degree a homogeneous liquid
characterized by the bulk thermodynamic properties. This homogeneous
density plays a role of the collective order parameter. Therefore
the ice nucleation suppression in nanopores is not a microscopic,
but rather a macroscopic effect. The inhomogeneous water density profile
observed in the neutron scattering experiments does take place but
apparently does not influence observable thermodynamic characteristics.
These arguments have a general character and hold, for example, for
melting of nm-sized crystalline solid particles \cite{couchman1977thermodynamic,mei2007melting}
and the derivation of Kelvin equation for the pressure of vapor in
equilibrium with liquid droplets \cite{LandauLifshitzStatPhys5}.
In each of the cases the thermodynamics properties of the infinite
(bulk) medium were successfully used to describe the properties of
nm-sized particles. The similarity makes us expect that freezing phase
transition of confined water in the pores should be smeared-out in
the PT plane due to essential fluctuations related to the finite number
of molecules in pores.

We presented a line of arguments to support the collective nature
of the phase transitions and a deep relation between the bulk and
confined liquid properties as soon as the water samples contained
in nanopores of sufficiently large radii, $R\gg t$. One of the recent
examples concerns the measurements of the dielectric properties of
liquid water confined in MCM-41 nanopores of $D=3.5$nm size and thus
prevented from freezing well below the natural freezing point at temperatures
roughly corresponding to the so called $\lambda-$point first observed
in the bulk water\cite{hodge1978relative}. The dielectric constant
of the water samples manifests a profound bump in the dielectric constant
rising from the typical value $\epsilon\sim10^{2}$ to as much as
$\epsilon\sim2\cdot10^{4}$ within the narrow temperature range $\triangle T\sim1K$
near $T_{F}^{exp}\approx235K$ in the liquid phase \cite{fedichev2011experimental,bordonskiy2012study}
in accordance with the earlier theoretical predictions \cite{men2011possible}.
Since the dielectric susceptibility increase is a signature of a ferroelectric
phase \cite{frohlich1949theory} and the ferroelectric ordering is
a collective phenomenon, we can argue that ferroelectric features
observed in nanopores do occur and should be eventually observed in
the bulk liquid water near the $\lambda-$point. 

\bibliographystyle{apsrev4-1}
\bibliography{../../Qrefs}

\end{document}